 \documentclass[aps,twocolumn,superscriptaddress,showpacs]{revtex4}

\usepackage{epsf}
\usepackage{amsmath}
\usepackage{amssymb}

\newcommand{\ER}[1]{{E_{R}^{\rm (#1)}}}
\newcommand{\RHO}[1]{{\rho^{\rm (#1)}}}
\newcommand{\SIGMA}[1]{{\sigma^{(\rm #1)}}}

\def\<{\langle}
\def\>{\rangle}

\begin{document}

\title{Comparison of the relative entropy of entanglement and
negativity}

\author{Adam Miranowicz}
\affiliation{Faculty of Physics, Adam Mickiewicz University,
61-614 Pozna\'n, Poland}
\affiliation{Institute of Theoretical Physics and Astrophysics,
University of Gda\'nsk, 80-952 Gda\'nsk, Poland}

\author{Satoshi Ishizaka}
\affiliation{Nano Electronics Research Laboratories, NEC
Corporation, 34 Miyukigaoka, Tsukuba 305-8501, Japan}
\affiliation{INQIE, The University of Tokyo, 4-6-1 Komaba,
Meguro-ku, Tokyo 153-8505, Japan}

\author{Bohdan Horst}
\affiliation{Faculty of Physics, Adam Mickiewicz University,
61-614 Pozna\'n, Poland}

\author{Andrzej Grudka}
\affiliation{Faculty of Physics, Adam Mickiewicz University,
61-614 Pozna\'n, Poland} \affiliation{Institute of Theoretical
Physics and Astrophysics, University of Gda\'nsk, 80-952 Gda\'nsk,
Poland} \affiliation{National Quantum Information Centre of
Gda\'{n}sk, 81-824 Sopot, Poland}

\date{\today}

\begin{abstract}

It is well known that for two qubits the upper bounds of the
relative entropy of entanglement (REE) for a given concurrence as
well as the negativity for a given concurrence are reached by pure
states. We show that, by contrast, there are two-qubit mixed
states for which the REE for some range of a fixed negativity is
higher than that for pure states. Moreover, we demonstrate that a
mixture of a pure entangled state and pure separable state
orthogonal to it is likely to give the maximal REE. By noting that
the negativity is a measure of entanglement cost under operations
preserving positivity of partial transpose, our results provide an
explicit example of operations such that, even though the
entanglement cost for an exact preparation is the same, the
entanglement of distillation of a mixed state can exceed that of
pure states. This means that the entanglement manipulation via a
pure state can result in a larger entanglement loss than that via
a mixed state.
\\

\end{abstract}


\maketitle

\pagenumbering{arabic}

\section{Introduction}

In quantifying quantum entanglement of two-qubit mixed states,
various measures are commonly applied \cite{Horodecki-review}: the
relative entropy of entanglement (REE) \cite{Vedral97a} -- a
measure of the ``distance'' (or distinguishability) of an
entangled state from the set of disentangled states, the
(logarithmic) negativity \cite{Peres,Horodecki} -- a measure of
entanglement cost under operations preserving the positivity of
partial transpose (PPT) \cite{Audenaert}, and the concurrence
\cite{Wootters} -- a measure of the entanglement of formation
\cite{Bennett1}.

It can be shown analytically that the upper bounds of the REE for
a given concurrence \cite{Vedral98} and of the negativity for a
given concurrence \cite{Verstraete} are reached by pure states. So
one could conjecture that pure states have also the highest REE
for a given negativity. However, we will demonstrate that there
are mixed states exhibiting the REE for a given negativity (in
some range) higher than for pure states. Before going into details
let us briefly describe the entanglement measures.

\section{Entanglement measures}

The relative entropy of entanglement in two-qubit systems
according to Vedral {\em et al.} can be defined as
\cite{Vedral97a,Vedral98}
\begin{equation}
E_R(\rho)={\rm min}_{\sigma' \in {\cal D}} S(\rho ||\sigma')
=S(\rho ||\sigma ), \label{N01}
\end{equation}
where the minimum is taken over the set ${\cal D}$ of all
separable states $\sigma$, and $S$ is the quantum relative entropy
\begin{equation}
S(\rho ||\sigma )={\rm Tr}\,( \rho \log_2 \rho -\rho\log_2 \sigma
) \label{N02}
\end{equation}
between states $\rho$ and $\sigma$. The REE measures a
quasidistance, say $D(\rho ||\sigma )$, of the entangled state
$\rho$ from the closest separable state (CSS) $\sigma$. Based on
the quantum version of Sanov's theorem, one can also interpret the
REE as a measure of statistical distinguishability of $\rho$. The
choice of $S(\rho ||\sigma )$ as a candidate for $D(\rho ||\sigma
)$ is by no means unique, although this is, to our knowledge, the
only proposal that coincides for pure states with the von Neumann
entropy of the reduced density operator. Also note that $S(\rho
||\sigma )$ is not symmetric and nor does it satisfy the triangle
inequality; thus it is not a true metric.

The negativity $N({\rho })$ for a two-qubit state $\rho$ can be
defined by \cite{Zyczkowski98,Eisert,Vidal}:
\begin{equation}
{N}({\rho })=\max \{0,-2\mu _{\min}\},  \label{N03}
\end{equation}
where $\mu _{\min}=\min{\rm eig}(\rho^{\Gamma})$ is the minimal
eigenvalue of the partial transpose, denoted by $\Gamma$, of
$\rho$. The negativity is directly related to the Peres-Horodecki
criterion \cite{Peres,Horodecki}. The logarithmic negativity,
given by $\log_2[{N} ({\rho})+1]$, is a measure of the
entanglement cost $E_C({\rho})$ under PPT operations
\cite{Audenaert,Ishizaka04}. The negativity and logarithmic
negativity are monotonically related, reaching unity for Bell
states and vanishing for separable states. So for simplicity of
our further analysis, we use the negativity instead of the
logarithmic negativity.

Another measure of entanglement is the entanglement of formation
$E_{F}({\rho})$ \cite{Bennett1} or, equivalently for two qubit
states, the Wootters concurrence \cite{Wootters} defined as
$C({\rho})=\max \{0,2\max_j\lambda_j-\sum_j\lambda_j\}$, where the
$\lambda _{j}$'s stand for the square roots of the eigenvalues of
${\rho }({\sigma }_{y}\otimes {\sigma }_{y}){\rho}^{\ast }({ \rho
}_{y}\otimes {\sigma }_{y})$, and ${\sigma }_{y}$ is the Pauli
spin matrix.

In the last section, we also analyze the entanglement of
distillation, $E_D({\rho})$ \cite{Bennett1}, a measure of the
entanglement as the fraction of Bell states that can be distilled
using the optimal purification protocol.

\section{REE with fixed $N$ for pure and mixed states}

\vspace{0mm}
\begin{figure}
 \epsfxsize=8.5cm\centerline{\epsfbox{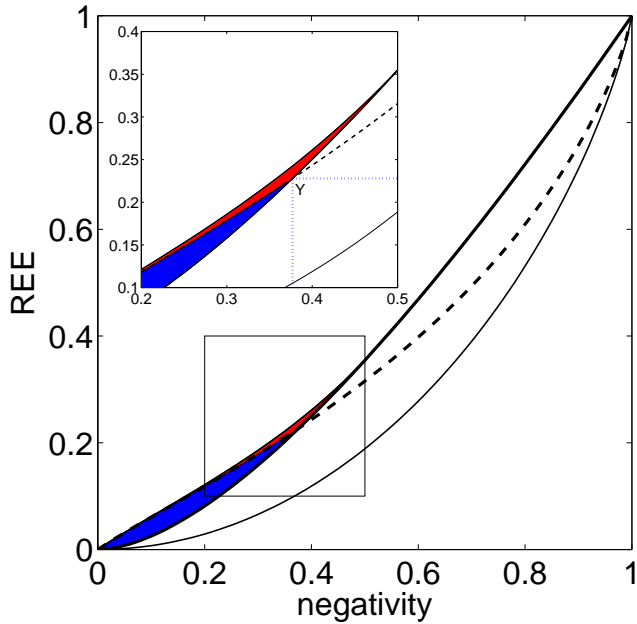}}
\caption{REE $E_R(\rho)$ as a function of negativity $N(\rho)$ for
pure states $\RHO{P}$ (thick solid curves), the (standard)
Horodecki states $\RHO{H}$ (dashed curves), the optimal
generalized Horodecki states $\RHO{OGH}$ (uppermost thin solid
curves), and the Bell-diagonal states $\RHO{BD}$ (lowest solid
curves). Marked regions correspond to states exceeding the
pure-state REE: Blue (red) region shows states $\RHO{H'}$
($\RHO{GH}$) for which $\ER{H'}>\ER{P}$
($\ER{GH}>\max\{\ER{P},\ER{H}\}$). }
\end{figure}

The REE and the entanglement of formation coincide for pure
states, but for mixed states the inequality $E_{F}({\rho})\ge
E_R({\rho})$ holds \cite{Vedral98}. As the concurrence is
monotonically related to the entanglement of formation for an
arbitrary state, the upper bound of the REE for a given
concurrence is reached for pure states. On the other hand, as
shown by Verstraete {\em et al.} \cite{Verstraete}, the negativity
$N({\rho})$ of an arbitrary state can never exceed its concurrence
$C({\rho})$. The upper bound of the negativity for a given
concurrence, i.e., $C({\rho})=N({\rho})$, is reached for a class
of states for which the eigenvector of the partially transposed
$\rho$ corresponding to the negative eigenvalue is a Bell state
\cite{Verstraete,Miran1,Miran2}. Pure states and also some mixed
states (including Bell-diagonal states) belong to this class.
Thus, we see that mixed states cannot give higher values of the
REE and negativity for a given concurrence than those for pure
states. In the following we will show that the mixed-state REE can
exceed the pure-state REE for a given negativity.

An arbitrary two-qubit pure state can be changed by local
rotations into a state of the form ($0\le P\le 1$)
\begin{equation}
|\psi_{P} \rangle =\sqrt{P} |01\rangle + \sqrt{1-P} |10\rangle,
\label{N04}
\end{equation}
as can be shown by applying the Schmidt decomposition
\cite{Peres-book}. The negativity of $|\psi_{P} \rangle$ is simply
described by $N(|\psi _{P}\rangle) =2\sqrt{P(1-P)}$, while the
REE, being equal to the entanglement of formation, can be given as
a function of $N\equiv N(|\psi _{P}\rangle)$ as
\begin{equation}
\ER{P}(N) \equiv E_R(|{\psi}_{P}\rangle) = H_2\left(
\frac{1}{2}[1+\sqrt{1-N^{2}}]\right), \label{N05}
\end{equation}
where $H_2(x)=-x\log_2 x-(1-x)\log_2 (1-x)$ is the binary entropy.
Equation (\ref{N05}) corresponds to the well-known Wootters
relation between the concurrence and the entanglement of formation
\cite{Wootters}, since
$N(|{\psi}_{P}\rangle)=C(|{\psi}_{P}\rangle)$ and
$E_R(|{\psi}_{P}\rangle)=E_{F}(|{\psi}_{P}\rangle)$.

In comparison with pure states, let us analyze a mixture of a
maximally entangled state, say the ``triplet'' state
$|\psi^{+}\rangle=(|01\rangle+|10\rangle)/\sqrt{2}$, and a
separable state orthogonal to it, say $|00\rangle$, i.e.,
\cite{Horodecki-book},
\begin{equation}
\RHO{H}=p|\psi^{+}\rangle \langle \psi^{+}|+(1-p)|00\rangle
\langle 00|, \label{N06}
\end{equation}
where the parameter $p\in\langle 0,1 \rangle$. For brevity, we
shall refer to (\ref{N06}) as the Horodecki state, although
alternatively it could be named after others (see, e.g.,
\cite{Vedral98,Verstraete}). The negativity of the Horodecki state
reads as
\begin{equation}
N(\RHO{H})=\sqrt{ (1-p)^{2}+p^{2}}-(1-p), \label{N07}
\end{equation}
while the REE as a function of $N\equiv N(\RHO{H})$ can be given
by Vedral-Plenio's formula \cite{Vedral98}
\begin{eqnarray}
\ER{H}(N) &\equiv& E_R(\RHO{H}) = 2H_2(1-p/2)-H_2(p)-p \quad \notag \\
&=& (p-2)\log_2 (1-p/2) +(1-p)\log_2 (1-p), \label{N08}
\end{eqnarray}
where $p=\sqrt{2N(1+N)}-N$. By comparing the REEs for the
Horodecki and pure states we observe that
\begin{subequations}
\begin{eqnarray}
 \ER{H}(N) > \ER{P}(N)& \quad &{\rm for}\; 0<N<N_Y,\label{N09a}
\\
 \ER{H}(N) < \ER{P}(N)& \quad &{\rm for}\; N_Y<N<1, \label{N09b}
\end{eqnarray}
\end{subequations}
where $N_Y=0.3770\ldots$ and $\ER{H}(N_Y) =
\ER{P}(N_Y)=0.2279\ldots$ as shown in the inset plot of Fig. 1.
The inequality (\ref{N09a}) can also be shown by expanding
(\ref{N05}) and (\ref{N08}) in power series of $N$ close to zero,
then one gets $\ER{H}(N)= N(1-\sqrt{N/2})/\ln 4+{\cal O}(N^2)>0$
and negligibly small $\ER{P}(N) \sim {\cal O}(N^2)$. To show
inequality (\ref{N09b}) more clearly, we can expand (\ref{N05})
and (\ref{N08}) around $N=1-\epsilon$ close to one, then we have
$\ER{P}(N) = 1 - \epsilon/\ln 2+ {\cal O}(\epsilon^2)$, which is
greater than $\ER{H}(N)= 1 - \epsilon (1 - \ln\epsilon)/\ln 4+
{\cal O}(\epsilon^2)$. Thus, a comparison of (\ref{N05}) and
(\ref{N08}) demonstrates the main point of this paper: There are
mixed states having the REE for a given negativity (in some range)
higher than that of pure states.

\begin{figure}[h]
\epsfxsize=9cm\centerline{\epsfbox{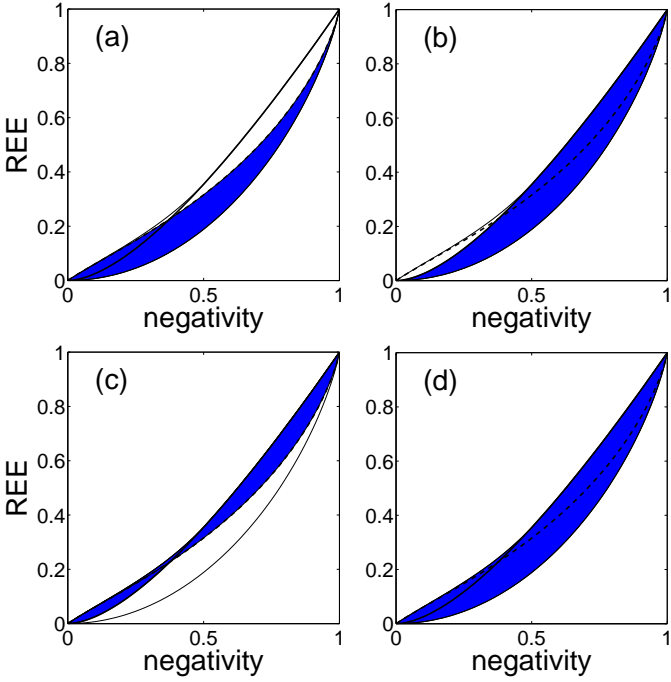}}
 \caption{Ranges of values of the REE with a given negativity
 for the following states:
 (a) $\RHO{H'}$ given by (\ref{N10}), (b) $\RHO{P'}$ given by
(\ref{N14}), (c) $\RHO{GH}$ given by (\ref{N19}), and (d)
$\RHO{GH'}$ given by (\ref{N26}). Curves are the same as in Fig.
1.}
\end{figure}

So far, we have analyzed the Horodecki states, which correspond to
the broken curve in Figs. 1 and 2. Now, we will give analytical
examples of mixed states more entangled than pure states, which
could correspond to any point in the dark region in Fig. 2(a). The
desired mixed states can be generated from the Horodecki state
$\RHO{H}$ by mixing it with the separable state $\SIGMA{H}$
closest to $\RHO{H}$ as follows:
\begin{eqnarray}
\RHO{H'}(p,N)= (1-x)\RHO{H}+x\SIGMA{H}, \label{N10}
\end{eqnarray}
defined for $N\in\langle 0,1 \rangle$ and
$p\in\langle\sqrt{2N(1+N)}-N,1\rangle$, where
\begin{eqnarray}
x=\frac{(N+p)^2-2N(1+N)}{p^2(1+N)} \label{N11}
\end{eqnarray}
and the corresponding CSS is given by ($q=p/2$)
\begin{eqnarray}
\SIGMA{H}(p) &=&
{q}(1-{q})\sum_{j,k=0}^1(-1)^{j-k}|j,1-j\rangle \langle k,1-k| \notag \\
&&+(1-{q})^{2}|00\rangle \langle 00|+{q}^{2}|11\rangle \langle 11|.
\label{N12}
\end{eqnarray}
By virtue of the Vedral-Plenio theorem \cite{Vedral98}, state
(\ref{N12}) is the CSS for $\RHO{H'}$ for any
$x\in\langle0,1\rangle$. Thus, we find that the REE for
$\RHO{H'}(p,N)$ is given by
\begin{eqnarray}
E_R(\RHO{H'}) \equiv \ER{H'}(p,N) \hspace{4.2cm}
\label{N13} \\
= q^2 x\log_2 x+2qy_1\log_2 \left(\frac{y_1}{1-q}\right) +
y_2\log_2 \left(\frac{y_2}{(1-q)^2}\right) \nonumber
\end{eqnarray}
where $y_1=1-qx$, $y_2=1-2q+q^2x$. The choice of $x$, given by
(\ref{N11}), implies that $N$ is just the negativity of
$\RHO{H'}(p,N)$. For $p=p_0\equiv\sqrt{2N(1+N)}-N$, the state
(\ref{N10}) goes into the Horodecki state, given by (\ref{N06}).
States corresponding to all points in the blue region in Fig. 1
can be generated from $\RHO{H'}(p,N)$ by changing $N$ from 0 to
$N_Y$ and slightly increasing $p$ from the value of $p_0$. By
choosing properly $N\in\langle 0,1 \rangle$ and $p\in\langle
p_0,1\rangle$, the state $\RHO{H'}(p,N)$ corresponding to any
point in the entire dark region in Fig. 2(a) can be generated. It
is seen that pure and mixed states having the negativity higher
than that $\RHO{H}$, which correspond to the white region under
the thick solid curve in Fig. 2(a), are not included in the family
of states $\RHO{H'}$. By contrast, dark region in Fig. 2(b)
corresponds to states that can be obtained from pure states
$|\psi_{P} \rangle$ by mixing them with the separable states
$\SIGMA{P}= P|01\rangle\langle01|+ (1-P)|10\rangle\langle10|$
closest to $\RHO{P}$. They can be given, in analogy to
(\ref{N10}), as
\begin{eqnarray}
\RHO{P'}(P,N)= (1-x)|\psi_{P} \rangle\langle \psi_{P} |
+x\SIGMA{P}, \label{N14}
\end{eqnarray}
where $x=1-N/[2\sqrt{P(1-P)}]$ for $N\in\langle 0,1\rangle$ and
$P\in\langle P_-,P_+\rangle$ with
$P_{\pm}=\frac12(1\pm\sqrt{1-N^2})$. The bounds on $P$ are
obtained from the requirement that $\RHO{P'}(P,N)$ should be a
positive semidefinite operator. In special cases for $P=P_{\pm}$,
the mixed state $\RHO{P'}(P_{\pm},N)$ becomes the pure state
$\RHO{P}(N)$. In analogy to the state (\ref{N10}), the
Vedral-Plenio theorem guarantees that the CSS for $\RHO{P'}(P,N)$
is the same as for the pure state $|\psi_{P} \rangle$. Thus, we
can calculate the REE for (\ref{N14}) arriving at
\begin{eqnarray}
E_R(\RHO{P'}) \equiv \ER{P'}(P,N) \hspace{4.2cm} \label{N15}\\
= H_2(P) -\frac{z-Py_{-}}{2P-y_{+}}\log_2
\Big(\frac{y_{-}}{2}\Big)
-\frac{z-Py_{+}}{2P-y_{-}}\log_2\Big(\frac{y_{+}}{2}\Big), \notag
\end{eqnarray}
where $y_{\pm}=1\pm\sqrt{1-2z}$ and
$z=2P(1-P)x(2-x)=2P(1-P)-N^2/2$. The REE, given by (\ref{N15}),
for $P=P_{\pm}$ goes into (\ref{N05}) as expected. The lower bound
of the REEs for both $\RHO{P'}(P,N)$ and $\RHO{H'}(p,N)$ is the
same and given by
\begin{eqnarray}
\ER{BD}(N)&\equiv& \ER{P'}(1/2,N)=\ER{H'}(1,N) \nonumber\\ &=&
1-H_2\Big(\frac{1+N}{2}\Big). \label{N16}
\end{eqnarray}
With the help of the Vedral {\em et al.} results \cite{Vedral97a},
we can conclude that the REE, given by (\ref{N16}), is the same as
for a Bell-diagonal state defined by
\begin{eqnarray}
\RHO{BD}=\sum_{i=0}^3 \lambda_i |\beta_{i} \rangle\langle
\beta_{i} |, \label{N17}
\end{eqnarray}
where $|\beta_{i} \rangle$ are the Bell states,
$\sum_j\lambda_j=1$, $\max_j\lambda_j=(1+N)/2>1/2$, and $N$ is the
negativity $N(\RHO{BD})$. Specifically, the states (\ref{N10}) for
$p=1$ and (\ref{N14}) for $p=1/2$ go into the following
Bell-diagonal states
\begin{equation}
\RHO{H'}(1,N)=\frac{1-N}{4}(|\beta_0\rangle
\langle\beta_0|+|\beta_2\rangle\langle\beta_2|)
+\frac{1+N}{2}|\beta_3\rangle \langle\beta_3|, \notag
\end{equation}
\begin{equation}
\RHO{P'}(1/2,N)= \frac{1+N}{2}|\beta_1\rangle\langle\beta_1|
+\frac{1-N}{2}|\beta_3\rangle\langle\beta_3|,
  \label{N18}
\end{equation}
respectively, where the Bell states are given in the following
order $|\beta_{2j+k} \rangle=[|0,k\rangle+(-1)^j
|1,1-k\rangle]/\sqrt{2}$.

The Horodecki state is more entangled than the pure state at most
at $N\equiv N'=0.1539\ldots$ for which $\max_N [\ER{H}(N) -
\ER{P}(N)]=0.0391\ldots$. The question arises about the highest
possible difference for an arbitrary mixed state. This problem is
strictly related to finding the upper bound of the REE vs
negativity.

\section{Numerical simulations}

There has been a long-standing open problem of finding a closed
analytical formula for the REE for two qubits, which corresponds
to finding the $\sigma$ for a given entangled state $\rho$
\cite{Eisert2} and it is argued that the analytical solution does
not exist \cite{Miran07}. Moreover, there has not yet been an
efficient numerical method proposed to calculate the REE for an
arbitrary entangled mixed state even in case of two qubits.
Analytical formulas for the REE are known only for some special
sets of states with high symmetry
\cite{Vedral97a,Vedral97b,Vedral98,Vollbrecht,Audenaert01,Audenaert02,Miran07}.
Thus, usually, numerical methods for calculating the REE have to
be applied \cite{Vedral98,Rehacek,Doherty}. The complexity of the
two-qubit problem can be explained by virtue of Caratheodory's
theorem, which implies that minimalization of the quantum relative
entropy $S(\rho ||\sigma )$ should be performed over $79$ real
parameters describing decomposed $\sigma$ \cite{Vedral98}. Usually
\cite{Vedral98,Rehacek}, gradient-type algorithms are applied to
perform the minimalization. \v{R}eh\'a\v{c}ek and Hradil
\cite{Rehacek} proposed a method resembling a state reconstruction
based on the maximum likelihood principle. Doherty {\em et al.}
\cite{Doherty} designed a hierarchy of more and more complex
operational separability criteria for which convex optimization
methods (known as semidefinite programs) can be applied
efficiently.

\begin{figure}
\epsfxsize=8cm\centerline{\epsfbox{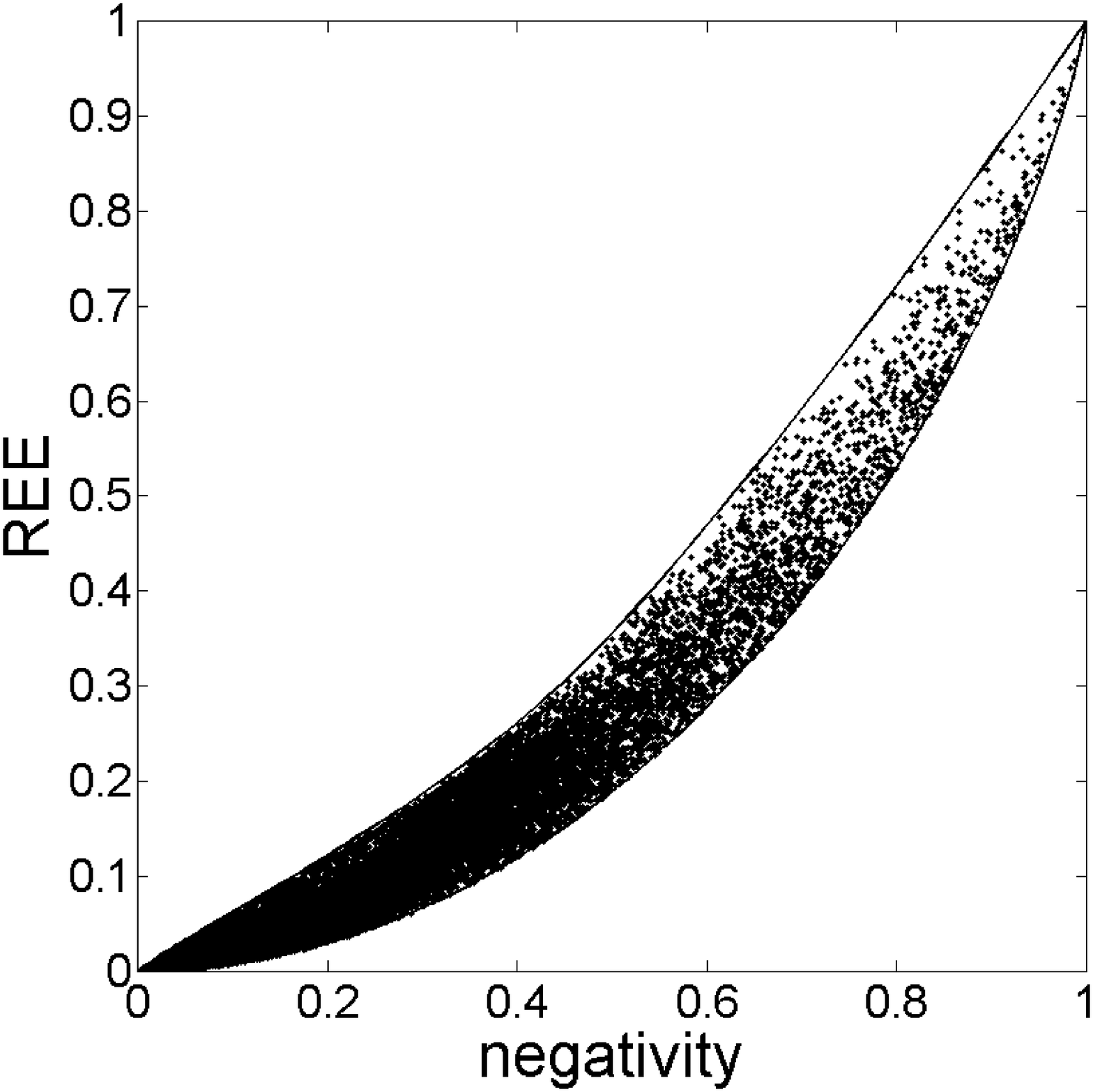}}
\epsfxsize=8cm\centerline{\epsfbox{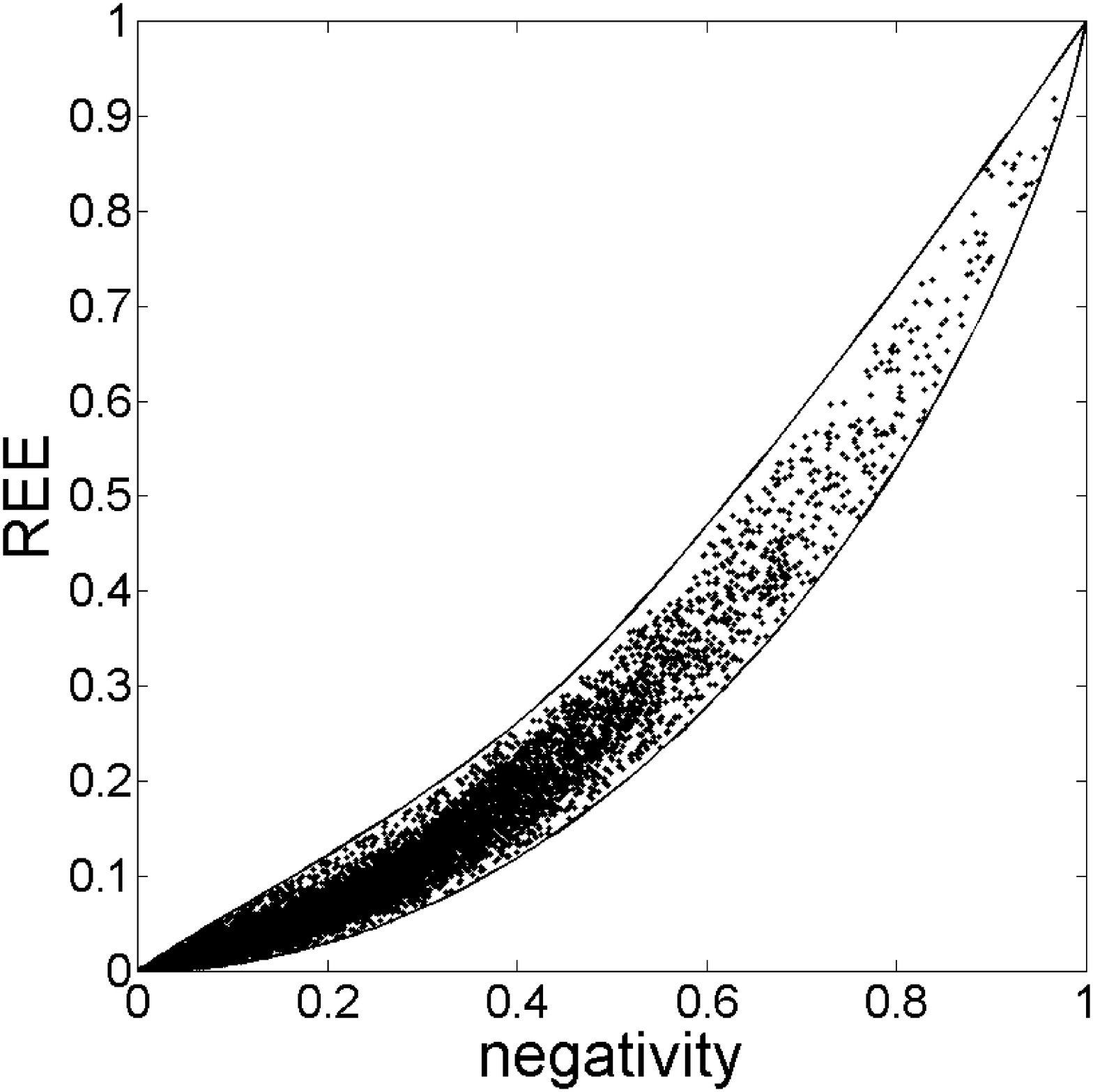}} \caption{REE
$E_R(\rho)$ versus negativity $N(\rho)$ for randomly-generated
states $\rho$ of rank 2 (upper figure) and rank 3 (lower figure).
The upper- and lowermost curves correspond to the optimal
generalized Horodecki states and Bell-diagonal states,
respectively. Random rank-4 states lie in the same range.}
\end{figure}

Nevertheless, there is a compact-form solution to the inverse
problem, which characterizes an entangled state $\rho$ for a given
full-rank $\sigma$ \cite{Ishizaka03,Miran07}:
\begin{eqnarray}
  \rho &=& \sigma - x G(\sigma),
  \label{N50}
\end{eqnarray}
\begin{eqnarray}
   G(\sigma) &=& \sum_{i,j} G_{ij}|i\rangle\langle i|
  (|\phi\rangle\langle \phi|)^\Gamma |j\rangle\langle j|,
  \label{N51}
\end{eqnarray}
and
\begin{equation}
G_{ij}\equiv\left\{
\begin{array}{cl}
\gamma_i &\hbox{~~for $\gamma_i\!=\!\gamma_j$} \cr
\frac{\gamma_i-\gamma_j}{\ln \gamma_i-\ln \gamma_j}
         &\hbox{~~for $\gamma_i\!\ne\!\gamma_j$,}
\end{array}
\right.  \label{N52}
\end{equation}
and $|\phi\rangle$ is the kernel of $\sigma^\Gamma$, while $|i>$
and $\gamma_i$ are eigenstates and eigenvalues of $\sigma$,
respectively. Thus, the REE is given by
\begin{eqnarray}
  E_R(\rho) = S(\sigma) - S(\rho) + x {\rm tr}
  \big[ (|\phi\rangle\langle \phi|)^\Gamma \sigma \log_2 \sigma
  \big],
  \label{N53}
\end{eqnarray}
where $S(\cdot)$ is the von Neumann entropy. In the following
$x_{\max}$ denotes the largest $x$ in (\ref{N50}), for which
$\rho$ is a positive operator. The solution can be applied also
for lower-rank CSSs in a limiting sequence from a full-rank state
by noting that the REE is a continuous function.

We have applied two approaches in our numerical simulations. In
the standard approach, the states are chosen at random and their
$E_R$ and $N$ are calculated numerically using a simplex search
method without using numerical or analytic gradients. However,
given the fact that no closed formula exists for $E_R$
\cite{Eisert2,Miran07}, it is more logical to choose random states
(call them $\sigma$) on the boundary of the separability region
and generate entangled states $\rho$ for which $\sigma$ is the CSS
by applying Eqs. (\ref{N50})--(\ref{N53}). The latter method is
faster by three orders of magnitude than the standard one. Figure
3 shows the results of our simulations for random rank-2 and
rank-3 states. For brevity, we have omitted a similar figure for
random rank-4 states. The simulations confirm our analytical
predictions that the mixed-state REE can exceed the pure-state REE
but also indicate lower and upper bounds of the REE vs negativity.

\section{REE for the generalized Horodecki states}

Our numerical simulations, partially shown in Fig. 3, indicate
that the upper bound $E_R(N)$ can be given by the rank-2
generalized Horodecki state $\RHO{GH}$ defined as follows
\cite{Miran07}:
\begin{eqnarray}
  \RHO{GH} &=& p |\psi_P\> \< \psi_P| + (1-p)|00\>\<00|,
\label{N19}
\end{eqnarray}
where $|\psi_P\>$ is given by (\ref{N04}) and $p,P\in \langle 0,1
\rangle$. In the special case of $P=1/2$, $\RHO{GH}$ reduces to
the standard Horodecki state, while $\RHO{GH}$ for $p=1$
corresponds to a pure state $|\psi_P\>$. Note that the negativity
$N \equiv N(\RHO{GH})$ is simply given by:
\begin{eqnarray}
  N =\sqrt{(1-p)^2+4p^2P(1-P)}-(1-p).
\label{N20}
\end{eqnarray}
By the inversion of this equation,
\begin{eqnarray}
  P &=& \frac{1}{2p}\left[p\pm \sqrt{p^2-N^2-2N(1-p)}\right],
\label{N21}
\end{eqnarray}
one can express the state $\RHO{GH}\equiv \RHO{GH}(p,N)$, given by
(\ref{N19}), as a function of negativity $N$ and parameter $p\ge
p_0(N)=\sqrt{2N(1+N)}-N$. The state $\RHO{GH}$ is a special case
of a more general state \cite{Miran07}
\begin{eqnarray}
\RHO{Z}\equiv \RHO{Z}\!\!\!\!\!\!\!_{x_{\max}}=
\begin{pmatrix}
r_1 & 0   & 0   & 0 \\
  0 & r_2 & y   & 0 \\
  0 & y   & r_3 & 0 \\
  0 & 0   & 0   & 0
\end{pmatrix},
\label{N22}
\end{eqnarray}
for which the CSS is given by:
\begin{eqnarray}
\SIGMA{Z}=
\begin{pmatrix}
R_1 & 0   & 0   & 0 \\
  0 & R_2 & Y   & 0 \\
  0 & Y   & R_3 & 0 \\
  0 & 0   & 0   & R_4
\end{pmatrix},
\label{N23}
\end{eqnarray}
where $Y=\sqrt{R_1R_4}$. Clearly, by assuming $y=\sqrt{r_2r_3}$,
the state $\RHO{Z}$ is reduced into the generalized Horodecki
state $\RHO{GH}$ with $r_1=1-p$, $r_2=Pp$ and $r_3=(1-P)p$. States
$\RHO{Z}$ and $\SIGMA{Z}$ are related by the following relations
assuming for convenience that $R_1\ge R_4$ \cite{Miran07}:
\begin{equation}
  r_2=R_2+\frac{2R_4}{z^{2}}(R_2^2-R_2R_3+2Y^2)+\frac{2R_4}{Lz}(R_2-R_3)
\label{N24}
\end{equation}
together with $r_1 = R_1-R_4$, $r_3=1-r_1-r_2$, and
$y=-[(r_2-R_2)(R_2-R_3)-2(R_1+R_2)R_4]/(2Y)$ given in terms of the
auxiliary functions $z=\sqrt{(R_2-R_3)^2+4Y^2}$ and
$L=\ln(R_2+R_3-z)-\ln(R_2+R_3+z)$. Moreover,
$x_{\max}=(R_1+R_4)/R_1$ if the condition $y=\sqrt{r_2r_3}$ is
satisfied for a given choice of $\{R_i\}$.

These equations can easily be inverted for $P=1/2$, which leads to
the solution given by (\ref{N12}) for the standard Horodecki state
$\RHO{H}$. By contrast, due to presence of logarithmic functions
of nonlinear combinations $\{R_i\}$ in the equations for
$\{r_i\}$, it looks impossible to invert the equations in order to
express all $\{R_i\}$ in terms of $\{r_i\}$ for the generalized
Horodecki state if $P\neq {0,\frac12,1}$. Thus, we can only give a
formula for the REE for $\RHO{GH}$ with $\{r_i\}$ as a function of
$\{R_i\}$:
\begin{equation}
  \ER{GH} = -H_2(r_1)-r_1 \log_2 R_1-f_-^2 \log_2 \lambda_- -f_+^2 \log_2 \lambda_+,
\label{N25}
\end{equation}
where $f_{\pm}={\cal N}_{\pm} [(\lambda_{\pm}-R_3)
\sqrt{r_2}+Y\sqrt{r_3}]$, $\lambda_{\pm}=\frac12(R_2+R_3\pm z)$
and ${\cal N}_{\pm}=[(\lambda_{\pm}-R_3)^2+Y^2]^{-1/2}$.

In any case, a multivariable numerical procedure for finding the
CSS $\SIGMA{GH}$ can be reduced to a single-variable problem;
namely, we can express $R_i$ (for $i=2,3,4$) in terms of $r_1$ and
$R_1$ as follows:
\begin{eqnarray}
  R_{2} &=& \frac14 (1+3r_1+2r_2-4R_1 -\sqrt{\delta}),
\nonumber \\
  R_{4} &=& R_1-r_1,
\nonumber \\
  R_{3} &=& 1-\sum_{i\ne 3} R_i, \label{N28}
\end{eqnarray}
where
\begin{eqnarray}
  r_{2,3} &=& \frac 12 \left[1-r_1\mp
  \sqrt{(1-r_1)^2-N(N+2r_1)}\right],
\nonumber \\
 \delta&=&(3r_1+1)^2-4r_2r_3-8R_1(r_1+1)
\nonumber \\
 &&+16\sqrt{R_1(R_1-r_1)r_2 r_3}. \label{N29}
\end{eqnarray}
Thus, to completely determine $\SIGMA{GH}$ for a given $\RHO{GH}$,
it is enough to find $R_1$ ($\ge r_1$) by numerically solving the
single-variable equation (\ref{N24}) with all the other variables
defined above.

A related problem is to find the optimal generalized Horodecki
state $\RHO{OGH}$, defined as $\RHO{GH}$ for a given $N$ and such
$p$, denoted by $p_{\rm opt}(N)$, for which the REE is maximized:
\begin{eqnarray}
\ER{OGH}(N) &\equiv& E_{R}[\RHO{OGH}(N)]
\nonumber \\
&\equiv& E_{R}[\RHO{GH}(p_{\rm opt}(N),N)]
\nonumber \\
&= & \max_{p\ge p_0(N)} E_{R}[\RHO{GH}(p(N),N)]. \label{N27}
\end{eqnarray}
The parameter $p_{\rm opt}$ can be found numerically by the
procedure described above. On the other hand, we have found a
fairly good approximation of $p_{\rm opt}$ for $0\le N\le 0.527$
given by
\begin{eqnarray}
  \bar p_{\rm opt}(N)= \frac13+\frac85 N -\frac7{11}N^2 \label{N30}
\end{eqnarray}
such that $E_{R}[\RHO{GH}(\bar p_{\rm opt}(N),N)]$ deviates by the
order $10^{-5}$ from the precise value of $\ER{OGH}(N)$. We also
find that $p_{\rm opt}$ becomes 1, so the optimum generalized
Horodecki state becomes a pure state $|\psi_P\>$ for $N {_> \atop
^\sim} 0.53$. It is worth noting that the precision of our
numerical calculations of the REE is $\sim 10^{-10}\div 10^{-8}$,
and $\max(\ER{H},\ER{P})$ is smaller than $\ER{GH}$ up to 0.0148
(at $N=0.377$), so it can clearly be distinguished from the
numerical noise.

The REE for the generalized Horodecki states $\RHO{GH}$ as a
function of $N$ for arbitrary values of $p$ correspond to the dark
region in Fig. 2(c). In analogy with the states $\RHO{H'}$, given
by (\ref{N10}), one can also define a class of more general states
by mixing $\RHO{GH}$ with its CSS $\SIGMA{GH}$, given by
(\ref{N12}), as follows:
\begin{eqnarray}
\RHO{GH'}= (1-x)\RHO{GH}+x\SIGMA{GH}, \label{N26}
\end{eqnarray}
where $x\in\<0,1\>$. As is seen in Fig. 2(d) in comparison to
Figs. 3(a) and 3(b), the REE vs $N$ for $\RHO{GH'}$ covers to
whole region of the values for randomly generated states.

We conjecture that for any two-qubit state $\rho$ described by the
REE $\ER{\rho}(N)\equiv E_R(\rho)$ as a function of the negativity
$N=N(\rho)$, the following inequalities are satisfied:
\begin{equation}
 \ER{OGH}(N) \ge \ER{\rho}(N) \ge \ER{BD}(N), \label{N31}
\end{equation}
which simplify to
\begin{equation}
 \ER{P}(N) \ge \ER{\rho}(N) \ge \ER{BD}(N)  \label{N32}
\end{equation}
for $N {_> \atop ^\sim} 0.53$, where $\ER{P}(N)$ and $\ER{BD}(N)$
are given by (\ref{N05}) and (\ref{N16}), while $\ER{GH}(N)$ is
found numerically by the described method using Eqs.
(\ref{N25})--(\ref{N29}). As a partial analytical support of our
conjectures, the extremal conditions for the REE with a fixed $N$
for $\RHO{GH}$ and $\RHO{BD}$ are examined in the next section. We
have also performed a numerical analysis, as discussed in Sec. VI,
to provide another support of validity of the conjecture. We have
generated altogether a few million random states $\rho$ of a fixed
rank (2, 3, and 4) and calculated the negativity and REE for each
of them.

\section{Some extremal conditions for REE with fixed $N$}

In the following, we show analytically that the Bell-diagonal
states and the generalized Horodecki states, thus also pure states
and the standard Horodecki states, satisfy some extremal
conditions for the REE with a fixed $N$ implied by a
Lagrange-multiplier method. Since negativity for a given state
$\rho$ is given by
\begin{eqnarray}
N(\rho)&=&-2 \min_{|\psi'\rangle} \big[
\langle\psi'|\rho^\Gamma|\psi'\rangle \big]
\label{N33} \\
&=&-2 \min_{|\psi'\rangle} \big[ \hbox{tr}\rho
(|\psi'\rangle\langle\psi'|)^\Gamma \big] \equiv -2 \big[
\hbox{tr}\rho (|\psi\rangle\langle\psi|)^\Gamma \big],
\nonumber
\end{eqnarray}
where $|\psi\rangle$ is the optimal state, let us consider the
following Lagrange function:
\begin{equation}
{\cal L}=\hbox{tr}\rho \log_2 \rho - \hbox{tr} \rho \log_2 \sigma
+ l \left(\hbox{tr} \rho
(|\psi\rangle\langle\psi|)^\Gamma+\frac{N}{2}\right), \label{N34}
\end{equation}
where $l$ is a Lagrange multiplier. For a small deviation of
\begin{equation}
\rho \rightarrow \rho+\Delta-(\hbox{tr}\Delta)\rho, \label{N35}
\end{equation}
where $\Delta$ is an arbitrary (but small) operator on the support
space of $\rho$ [denoted by $\hbox{supp}(\rho)$ hereafter], we
have
\begin{equation}
{\cal L} \rightarrow {\cal L} + \hbox{tr}\Delta \big[
\log_2\rho-\log_2\sigma+l(|\psi\rangle\langle\psi|)^\Gamma
-E_R(\rho)+\frac{l}{2}N(\rho)\big]. \label{N36}
\end{equation}
Since $\Delta$ is an arbitrary operator on $\hbox{supp}(\rho)$,
the following extremal condition is obtained:
\begin{equation}
P\big[\log_2\rho-\log_2\sigma+l(|\psi\rangle\langle\psi|)^\Gamma
-E_R(\rho)+\frac{l}{2}N(\rho)\big] P =0, \label{N37}
\end{equation}
where $P$ is the projector to $\hbox{supp}(\rho)$. Moreover, the
extremal condition for ${\cal L}$ with respect to $|\psi\rangle$
leads to the extremal condition for negativity, and thus
$|\psi\rangle$ must be the eigenstate corresponding to a negative
eigenvalue of $\rho^\Gamma$. Therefore, it is found that $\rho$,
its closest separable state $\sigma$, and the eigenstate
$|\psi\rangle$ corresponding to a negative eigenvalue of
$\rho^\Gamma$ should satisfy Eq.\ (\ref{N37}).

Now let us consider the case where $\rho$ is a mixed state of rank
2, i.e., $\rho=\lambda_1|e_1\rangle\langle e_1|+
\lambda_2|e_2\rangle\langle e_2|$, where $\{\lambda_i\}$ are
nonzero eigenvalues of $\rho$, and $|e_i\rangle$ are corresponding
eigenstates. The projector $P$ is then $P=|e_1\rangle\langle
e_1|+|e_2\rangle\langle e_2|$, and as a result the extremal
condition of Eq.\ (\ref{N37}) becomes
\begin{equation}
\langle e_1|\log_2 \sigma|e_2\rangle = l \langle
e_1|(|\psi\rangle\langle\psi|)^\Gamma|e_2\rangle, \label{N38}
\end{equation}
and
\begin{eqnarray}
\log_2\lambda_1 -\langle e_1|\log_2 \sigma|e_1\rangle + l \langle
e_1|(|\psi\rangle\langle\psi|)^\Gamma|e_1\rangle
\nonumber \\
=E_R(\rho)-\frac{l}{2}N(\rho), \label{N39} \\
\log_2\lambda_2 -\langle e_2|\log_2 \sigma|e_2\rangle + l \langle
e_2|(|\psi\rangle\langle\psi|)^\Gamma|e_2\rangle
\nonumber \\
=E_R(\rho)-\frac{l}{2}N(\rho). \label{N40}
\end{eqnarray}
However, Eqs. (\ref{N39}) and (\ref{N40}) are not independent of
each other. Indeed, for $\lambda_1\ne0$ and $\lambda_2\ne0$, these
equations are equivalent to
\begin{eqnarray}
\lambda_1\log_2\lambda_1 -\lambda_1\langle e_1|\log_2
\sigma|e_1\rangle + \lambda_1 l \langle
e_1|(|\psi\rangle\langle\psi|)^\Gamma|e_1\rangle
\nonumber \\
=\lambda_1 E_R(\rho)-\lambda_1\frac{l}{2}N(\rho), \label{N41} \\
\lambda_2\log_2\lambda_2 -\lambda_2\langle e_2|\log_2
\sigma|e_2\rangle + \lambda_2l \langle
e_2|(|\psi\rangle\langle\psi|)^\Gamma|e_2\rangle
\nonumber \\
=\lambda_2E_R(\rho)-\lambda_2\frac{l}{2}N(\rho), \label{N42}
\end{eqnarray}
and it is found that the sum of these equations is automatically
satisfied. Therefore, the extremal conditions for rank-2 states
are Eqs. (\ref{N38}) and (\ref{N39}) [or Eqs. (\ref{N38}) and
(\ref{N40})].

\subsection{Bell-diagonal states}

For the rank-2 Bell-diagonal states,
\begin{equation}
[\RHO{BD},\SIGMA{{BD}}]=[\RHO{BD},(|\psi\rangle\langle\psi|)^\Gamma]=0,
\label{N43}
\end{equation}
and hence $\langle e_1|\log_2\sigma|e_2\rangle= \langle
e_1|(|\psi\rangle\langle\psi|)^\Gamma|e_2\rangle=0$ holds again.
Equation (\ref{N38}) is then satisfied for any $l$, and the
extremal conditions are satisfied.

\subsection{Horodecki states}

For the standard Horodecki state, defined by (\ref{N06}), with its
CSS given by (\ref{N12}), we have:
\begin{eqnarray}
(\RHO{H})^\Gamma&=&q \big(|\psi^+\rangle\langle\psi^+|
+|\psi^-\rangle\langle\psi^-| +|00\rangle\langle11|
 \nonumber \\
&& ~~+|11\rangle\langle00|\big) + (1-p) |00\rangle\langle00|,
\label{N44}
\end{eqnarray}
and
\begin{eqnarray}
|\psi\rangle&=& \sqrt{s_-}|00\rangle -\sqrt{s_+}|11\rangle,
\nonumber \\
(|\psi\rangle\langle\psi|)^\Gamma&=& s_-|00\rangle\langle 00| +
s_+|11\rangle\langle 11| \cr
&&+\frac{tp}{2}(|\psi^-\rangle\langle\psi^-|-
|\psi^+\rangle\langle\psi^+|), \label{N45}
\end{eqnarray}
where $|\psi^{\pm}\rangle=(|01\rangle\pm|10\rangle)/\sqrt{2}$,
$s_{\pm}=[1\pm t(1-p)]/2$ and $t=1/\sqrt{2p^2-2p+1}$.

Since $\langle e_1|\log_2\sigma|e_2\rangle= \langle
e_1|(|\psi\rangle\langle\psi|)^\Gamma|e_2\rangle=0$, Eq.
(\ref{N38}) is satisfied for any $l$. Since there is only one
relation of Eq. (\ref{N39}) for $l$, the extremal conditions are
necessarily satisfied.

\subsection{Generalized Horodecki states}

Here, we show that the generalized Horodecki states are extremal.
The point is that only two extremal conditions should be satisfied
for the states: (i) given by (\ref{N38}) and (ii) given by either
of Eqs. (\ref{N39})--(\ref{N42}) or, e.g., the difference of Eqs.
(\ref{N41}) and (\ref{N42}). Condition (ii) is a linear function
of $l$, so it can easily be solved for $l$. The question is
whether the found $l$ also satisfies condition (i) or whether
left- and right-hand sides (LHS and RHS) of (i) are equal to zero.
In the following we show that the latter case is satisfied for the
generalized Horodecki states $\RHO{GH}$, and thus also for the
optimal states $\RHO{OGH}$, the standard Horodecki states
$\RHO{H}$, and pure states $\RHO{P}$.

For simplicity, we use the notation of Eqs. (\ref{N22}) and
(\ref{N23}) with the condition $y^2=r_2r_3$, which guarantees that
$\RHO{Z}$ and $\SIGMA{Z}$  become $\RHO{GH}$ and $\SIGMA{GH}$,
respectively. One finds that
\begin{eqnarray}
 |\psi\rangle &=& {\cal N} (-g |00\>+ 2y|11\>),
\nonumber \\
 (|\psi\rangle\langle\psi|)^\Gamma &=& {\cal N}^2
 \big[g^2 |00\>\<00|+4y^2|11\>\<11|
\nonumber \\
 &&\quad -2gy (|01\>\<10|+|10\>\<01|)\big],
\label{N46}
\end{eqnarray}
where $g=\sqrt{r_1^2+4y^2}-r_1$ and ${\cal N}=1/\sqrt{g^2+4y^2}$.
On the other hand, $\log_2\SIGMA{GH}$ can be calculated through
the eigenvalue decomposition
\begin{eqnarray}
  \SIGMA{GH} &=& R_1 |00\>\<00| + R_4 |11\>\<11|
\nonumber \\
  &&+ \lambda_+ |\lambda_+\>\<\lambda_+|
  + \lambda_- |\lambda_-\>\<\lambda_-|,
\label{N47}
\end{eqnarray}
where
\begin{eqnarray}
\lambda_{\pm} &=& \frac12\left[R_2+R_3\pm
\sqrt{(R_2-R_3)^2+4Y^2}\right],
\nonumber \\
|\lambda_{\pm}\>&=&{\cal
N}_{\pm}[(\lambda_{\pm}-R_3)|01\>+Y|10\>], \label{N48}
\end{eqnarray}
with ${\cal N}_{\pm}=[(\lambda_{\pm}-R_3)^2+Y^2]^{-1/2}$.
Moreover, for the nonzero eigenvalues, the eigenvectors of
$\RHO{GH}$ are found to be $|e_1\>=|00\>$ and
$|e_2\>=(1/\sqrt{y^2+r_3^2}) (y|01\>+r_3|10\>)$. Thus, it is seen
that both LHS and RHS of condition (\ref{N38}) are equal to zero.
The second condition is satisfied by choosing
\begin{eqnarray}
  l &=& \frac{2f(\ER{GH}-\log_2 r_1+\log_2 R_1)}{(f+1)(f-r_1)},
\label{N49}
\end{eqnarray}
where $f=\sqrt{r_1^2+4y^2}$. In any case, even without knowing
explicitly $\ER{GH}$ and $\{R_i\}$ in terms of $\{r_i\}$, we have
showed that the generalized Horodecki states satisfy the extremal
conditions.

\section{Conclusions}

In conclusion, we have demonstrated that there are mixed states
that have the REE in some range of a fixed negativity higher than
the pure-state REE for the same negativity. This is somewhat
surprising, since mixed states can neither exhibit the REE for a
given concurrence nor negativity for a given concurrence higher
than those for pure states. By applying Lagrange multipliers, we
have also shown that the Bell-diagonal states, pure states, but
also the so-called generalized Horodecki states, which are
mixtures of a pure entangled state and pure separable state
orthogonal to it, satisfy some extremal conditions for the REE
with a fixed negativity.

Our findings implicitly show another fact. For a given negativity,
the entanglement of distillation (ED) $E_D$ of the Horodecki state
$\RHO{H}$ can be larger than that of a pure state $\RHO{P}$. For
example, $N=0.1$ and $p=0.37$, one gets $E_D(\RHO{P})
=E_R(\RHO{P}) =0.025$ and $E_D(\RHO{H})>0.034$. Here, the lower
bound of the ED is given by $p^2/4$ (via the direct method shown
in \cite{Bennett1}). This lower bound can be slightly improved by
a method discussed in Ref. \cite{Ishizaka-notes}. The point is
that the logarithmic negativity is equal to a PPT entanglement
cost for an exact preparation, the REE is equal to a PPT
distillable entanglement for pure states, and the ED is a lower
bound of a PPT distillable entanglement. So our findings provide
an explicit example of PPT operations where, even though the
entanglement cost for an exact preparation is the same, the ED of
a mixed state can exceed that of pure states. In other words, the
entanglement manipulation via a pure state can result in a larger
entanglement loss than that via a mixed state.

{\bf Acknowledgments}. We thank Ryszard~Horodecki,\linebreak
Pa\-we\l{}~Horo\-decki, Micha\l{}~Horo\-decki, Zden\v{e}k Hradil,
Frank Verstraete, Shashank Virmani and Karol \.Zycz\-kowski for
their pertinent comments. We also appreciate useful suggestions of
an anonymous referee. The research was conducted within the LFPPI
network.

\end{document}